\documentclass[showpacs,twocolumn,preprintnumbers,amsmath,amssymb]{revtex4}

\usepackage{graphicx}
\usepackage{dcolumn}
\usepackage{bm}

\newcommand{\bq}{\begin{equation}}
\newcommand{\eq}{\end{equation}}
\newcommand{\bqa}{\begin{eqnarray}}
\newcommand{\eqa}{\end{eqnarray}}
\newcommand{\nn}{\nonumber \\}

\def\be     {\begin{equation}}
\def\ee     {\end{equation}}
\def\bea        {\begin{eqnarray}}
\def\eea        {\end{eqnarray}}
\def\bnn    {\begin{eqnarray*}}
\def\enn    {\end{eqnarray*}}

\begin{document}

\title{Emergent geometry from field theory: Wilson's renormalization group revisited}
\author{Ki-Seok Kim$^{1}$ and Chanyong Park$^{1,2}$}
\affiliation{$^{1}$Department of Physics, POSTECH, Pohang, Gyeongbuk 790-784, Korea \\ $^{2}$Asia Pacific Center for Theoretical Physics (APCTP), POSTECH, Pohang, Gyeongbuk 790-784, Korea}

\date{\today}

\begin{abstract}
We find a geometrical description from a field theoretical setup based on Wilson's renormalization group in real space. We show that renormalization group equations of coupling parameters encode the metric structure of an emergent curved space, regarded to be an Einstein equation for the emergent gravity. Self-consistent equations of local order-parameter fields with an emergent metric turn out to describe low energy dynamics of a strongly coupled field theory, analogous to the Maxwell equation of the Einstein-Maxwell theory in the AdS$_{d+2}$/CFT$_{d+1}$ duality conjecture. We claim that the AdS$_{3}$/CFT$_{2}$ duality may be interpreted as Landau-Ginzburg theory combined with Wilson's renormalization group, which introduces vertex corrections into the Landau-Ginzburg theory in the large$-N_{s}$ limit, where $N_{s}$ is the number of fermion flavors.
\end{abstract}


\maketitle

Hertz-Moriya-Millis theory has been a standard theoretical framework for non-Fermi liquid physics in metallic quantum criticality, which may be regarded to be essentially a mean-field theory \cite{QCP_Review}. However, recent developments doubt the validity of such a theoretical framework in a strong coupling regime \cite{Breakdown_Large_N}. Even if an artificial parameter is introduced to control the expansion, referred to as a flavor number of fermions, for example, spin degeneracy $N_{s}$, it turns out that a particular class of quantum corrections given by planar diagrams should be incorporated in the large$-N_{s}$ limit, resulting from the existence of abundant soft modes near a Fermi surface. Simply speaking, we do not have any theoretical paradigm for the so called Fermi-surface problem.

AdS$_{d+2}$/CFT$_{d+1}$ duality conjecture has been applied to the problem of quantum criticality \cite{AdS_CMP_Review}, hoping that this machinery takes into account such higher-order quantum corrections systematically beyond the present field theoretical framework \cite{AdS_CFT_Review}. In particular, symmetry properties and strong correlations may protect the universality class of such an AdS-liquid state, where fast thermalization is responsible for universal hydrodynamic transport phenomena \cite{AdS_Hydrodynamics}. Although the AdS$_{d+2}$/CFT$_{d+1}$ duality is expected to be Landau-Ginzburg-Wilson theory for strongly coupled field theories in one-dimensional higher curved space, it is desirable to reveal the clear connection between the field theoretical aspect and the geometrical description. This situation reminds us of the relationship between Landau Fermi-liquid theory and Landau-Ginzburg-Wilson theory for phase transitions in ``weakly" correlated electrons, clarified by BCS theory to reveal the connection between microscopic (fermions) and macroscopic variables (order parameter bosons) \cite{BCS_Review}.

In this study we find a geometrical description from a field theoretical setup based on Wilson's renormalization group \cite{Wilson_RG_Review} in real space. In particular, we show that the additional length scale to form a bulk geometry is identified with the step of the renormalization group procedure, where the limit of an infinite step corresponds to the infrared (IR) limit. As a result, renormalization group equations of coupling parameters encode the nature of an emergent curved space that their solutions give an emergent metric to describe. Taking into account the saddle-point approximation justified in the large$-N_{s}$ limit, we find self-consistent equations of local order-parameter fields in the emergent curved space, which describe low energy dynamics of a strongly coupled field theory. Comparing this framework with the Einstein-Maxwell theory in the AdS$_{d+2}$/CFT$_{d+1}$ duality conjecture \cite{AdS_CFT_Review}, we claim that renormalization group equations of coupling parameters and self-consistent equations of local order-parameter fields may be identified with the Einstein equation and the Maxwell equation in the gravity description, respectively. We demonstrate all of these aspects explicitly in $d = 1$, where the Wilson's renormalization group procedure in real space is well defined unambiguously. The AdS$_{3}$/CFT$_{2}$ duality may be interpreted as Landau-Ginzburg theory combined with Wilson's renormalization group, which introduces vertex corrections into the Landau-Ginzburg theory in the large$-N_{s}$ limit.

We focus on a spin chain, described by the Heisenberg model in $(1+1)D$. Resorting to the ``parton" construction \cite{Wen_Manybody_Textbook} of $\bm{S}_{i} = \frac{1}{2} f_{i\alpha}^{\dagger} \bm{\sigma}_{\alpha\beta} f_{i\beta}$, where $f_{i\sigma}$ is a fermion field with spin $\sigma$ at site $i$, we start from the path-integral representation of the Heisenberg spin chain \bqa && Z = \int D f_{i\sigma} \delta(f_{i\sigma}^{\dagger} f_{i\sigma} - N_{s} S) \exp\Bigl[- \int_{0}^{\beta} d \tau \Bigl\{ \sum_{i} f_{i\sigma}^{\dagger} \partial_{\tau} f_{i\sigma} \nn && + \frac{J}{N_{s}} \sum_{i} \bm{S}_{i} \cdot \bm{S}_{i+1} + g \sum_{i} \bm{H}_{i} \cdot \bm{S}_{i} \Bigr\} \Bigr] . \eqa Here, the constraint of $f_{i\sigma}^{\dagger} f_{i\sigma} = N_{s} S$ is introduced explicitly, where $N_{s}$ and $S$ denote the spin degeneracy and the size of spin, respectively, $N_{s} = 2$ and $S = 1/2$ in the case of the Heisenberg model. $J$ is an exchange coupling constant and $g$ is the Lande-$g$ factor. $\bm{H}_{i}$ is an applied magnetic field at site $i$, assumed to be $\bm{H}_{i} = H \bm{\hat{z}}$. $\beta = 1/T$ is an inverse temperature.

It is natural to decompose the Heisenberg interaction into particle-hole and particle-particle channels within the spin-singlet subspace, $\bm{S}_{i} \cdot \bm{S}_{i+1} = - \frac{1}{4} \Big( \hat{\chi}_{i i+1}^{\dagger} \hat{\chi}_{i i+1} + \hat{\Delta}_{i i+1}^{\dagger} \hat{\Delta}_{i i+1} \Big)$, where $\hat{\chi}_{i i+1} = f_{i+1 \sigma}^{\dagger} f_{i \sigma}$ measures the strength of exchange hopping and $\hat{\Delta}_{i i+1} = \varepsilon_{\sigma\sigma'} f_{i\sigma} f_{i+1\sigma'}$ with an antisymmetric tensor $\varepsilon_{\sigma\sigma'}$ represents that of singlet pairing \cite{Wen_Manybody_Textbook}. Although these two order parameters $\chi_{i i+1} = \langle \hat{\chi}_{i i+1} \rangle$ and $\Delta_{i i+1} = \langle \hat{\chi}_{i i+1} \rangle$ in the saddle-point approximation allow us to keep the particle-hole SU(2) gauge symmetry \cite{PH_SU2_Comment} explicitly, we do not take into account the pairing order parameter further just for simplicity. One can construct a geometrical description in the presence of the pairing order parameter, technically more involved. Later, we comment on this point.

Taking into account the uniform ansatz $\chi_{i i+1} \rightarrow \chi$, we obtain an effective lattice model, essentially ``noninteracting", given by \bqa && Z = \int D f_{i\sigma} \exp\Bigl[- \int_{0}^{\beta} d \tau \Bigl\{ \sum_{i} f_{i\sigma}^{\dagger} (\partial_{\tau} + \lambda + g \sigma H) f_{i\sigma} \nn && - \frac{J}{N_{s}} \chi \sum_{i} (f_{i\sigma}^{\dagger} f_{i+1\sigma} + f_{i+1\sigma}^{\dagger} f_{i\sigma}) \Bigr\} - \beta L \Big( z \frac{J}{N_{s}} \chi^{2} \nn && - N_{s} S \lambda \Big) \Bigr] , \eqa where $\lambda_{i} \rightarrow \lambda$ in the uniform ansatz is a Lagrange multiplier field to impose the single occupancy constraint, playing the role of a chemical potential in such fermions. $z$ is the coordination number, here $z = 2$.

The standard approach is to perform the saddle-point approximation, determining order parameters and revealing mean-field phases. Based on such a mean-field phase diagram, one can construct a low-energy effective field theory in a given mean-field phase and perform a perturbative renormalization group analysis to figure out a fixed-point structure near the mean-field phase beyond the mean-field description. Here, we suggest an approach beyond this framework, combining the mean-field theory with the Wilson's renormalization group framework.

The Wilson's renormalization group procedure in real space is as follows. First, we divide lattice fermions into two species: fermions on odd ($-$) sites and those on even ($+$) sites. Second, performing the path integral for fermion fields on odd sites, we obtain an effective theory for fermions on even sites. Third, we rescale the distance between even sites and accordingly, fermion fields also to make the effective theory turn back to the original expression. We repeat this procedure until reaching a fixed point. Suppose an effective theory in an $(n-1)^{th}$ step of the renormalization group procedure, given by \bqa && Z_{n} = 2^{(n-1) L} Z_{0} Z_{1} \cdot\cdot\cdot Z_{n-2} \int D \psi_{+i\sigma}^{(n-1)} D \psi_{-i\sigma}^{(n-1)} \nn && \times \exp\Bigl[- \sum_{i \omega} \Bigl\{ \sum_{i}' \psi_{+i\sigma}^{(n-1)\dagger} \mu_{n-1}(\chi,\lambda;i\omega;H) \psi_{+i\sigma}^{(n-1)} \nn && + \sum_{i}' \psi_{-i\sigma}^{(n-1)\dagger} \mu_{n-1}(\chi,\lambda;i\omega;H) \psi_{-i\sigma}^{(n-1)} \nn && - t_{n-1}(\chi,\lambda;i\omega;H) \sum_{i}' (\psi_{+i\sigma}^{(n-1)\dagger} \psi_{-i\sigma}^{(n-1)} \nn && + \psi_{-i\sigma}^{(n-1)\dagger} \psi_{+i+1\sigma}^{(n-1)} + H.c.) \Bigr\} - \beta L \Big( z \frac{J}{N_{s}} \chi^{2} - N_{s} S \lambda \Big) \Bigr] , \nn \eqa where the first step has been performed. $Z_{p} = \exp\Big( L N_{s} / 2 \sum_{i \omega} \ln \mu_{p}(\chi,\lambda;i\omega;H) \Big)$ is the partition function in the $p^{th}$ step. The factor of $2^{(n-1) L}$ appears as a result of rescaling. $\psi_{\pm i \sigma}^{(n-1)}$ is a lattice fermion at an even ($+$) or odd ($-$) sublattice of the site $i$ in an $(n-1)^{th}$ step. $\sum_{i}'$ means $\sum_{i = 1}^{L/2}$ instead of $\sum_{i = 1}^{L}$.

Performing the path integral for $\psi_{- i \sigma}^{(n-1)}$, we obtain an effective theory for $\psi_{+ i \sigma}^{(n-1)}$, given by \bqa && Z_{n} = 2^{(n-1) L} Z_{0} Z_{1} \cdot\cdot\cdot Z_{n-2} Z_{n-1} \nn && \int D \psi_{+i\sigma}^{(n-1)} \exp\Bigl[- \sum_{i \omega} \Bigl\{ \sum_{i}' \psi_{+i\sigma}^{(n-1)\dagger} \Big( \mu_{n-1}(\chi,\lambda;i\omega;H) \nn && - \frac{t_{n-1}^{2}(\chi,\lambda;i\omega;H)}{\mu_{n-1}(\chi,\lambda;i\omega;H)} \Big) \psi_{+i\sigma}^{(n-1)} - \frac{t_{n-1}^{2}(\chi,\lambda;i\omega;H)}{\mu_{n-1}(\chi,\lambda;i\omega;H)} \nn && \times \sum_{i}' \Bigl(\psi_{+i\sigma}^{(n-1)\dagger} \psi_{+i+1\sigma}^{(n-1)} + H.c. \Bigr) \Bigr\} - \beta L \Big( z \frac{J}{N_{s}} \chi^{2} - N_{s} S \lambda \Big) \Bigr] . \nn \eqa Rescaling the lattice and the fermion field accordingly, we rewrite this effective theory in an original form, given by \bqa && Z_{n} = 2^{n L} Z_{0} Z_{1} \cdot\cdot\cdot Z_{n-2} Z_{n-1} \nn && \int D \psi_{i\sigma}^{(n)} \exp\Bigl[- \sum_{i \omega} \Big\{ \sum_{i} \psi_{i\sigma}^{(n) \dagger} \mu_{n}(\chi,\lambda;i\omega;H) \psi_{i\sigma}^{(n)} \nn && - t_{n}(\chi,\lambda;i\omega;H) \sum_{i} \Bigl(\psi_{i\sigma}^{(n)\dagger} \psi_{i+1\sigma}^{(n)} + H.c. \Bigr) \Big\} \nn && - \beta L \Big( z \frac{J}{N_{s}} \chi^{2} - N_{s} S \lambda \Big) \Bigr] , \eqa where renormalized coupling parameters are given by \bqa && \mu_{n}(\chi,\lambda;i\omega;H) = \mu_{n-1}(\chi,\lambda;i\omega;H) - \frac{t_{n-1}^{2}(\chi,\lambda;i\omega;H)}{\mu_{n-1}(\chi,\lambda;i\omega;H)} , \nn && t_{n}(\chi,\lambda;i\omega;H) = \frac{t_{n-1}^{2}(\chi,\lambda;i\omega;H)}{\mu_{n-1}(\chi,\lambda;i\omega;H)} \eqa with $\mu_{0}(\chi,\lambda;i\omega;H) = - i\omega + \lambda + g \sigma H$ and $t_{0} = - \frac{J}{N_{s}} \chi$.

Performing the integration of $\psi_{i\sigma}^{(n)}$, we find an effective free energy in the $n^{th}$ step of the renormalization group procedure \bqa && F_{n}(\chi,\lambda;H) = - \frac{N_{s}/2}{\beta} \sum_{i\omega} \frac{1}{L} \sum_{k} \sum_{p = 0}^{n} \ln \Bigl\{ \mu_{p}(\chi,\lambda;i\omega;H) \nn && + \Theta(p - n + 1) z t_{p}(\chi,\lambda;i\omega;H) \gamma_{k} \Bigr\} + z \frac{J}{N_{s}} \chi^{2} - N_{s} S \lambda , \nn \eqa where the factor $1/2$ in $N_{s}/2$ is compensated by the $\sigma = \pm$ (spin) summation of the Zeeman term. Minimizing this effective free energy with respect to order parameters, we find self-consistent equations for such order parameters \begin{widetext}
\bqa && z \frac{J}{N_{s}} \chi = \frac{N_{s}/2}{\beta} \sum_{i\omega} \frac{1}{L} \sum_{k} \sum_{p = 0}^{n} \frac{\frac{\partial \mu_{p}(\chi,\lambda;i\omega;H)}{\partial \chi} + \Theta(p - n + 1) z \gamma_{k} \frac{\partial t_{p}(\chi,\lambda;i\omega;H)}{\partial \chi}}{ \mu_{p}(\chi,\lambda;i\omega;H) + \Theta(p - n + 1) z t_{p}(\chi,\lambda;i\omega;H) \gamma_{k} } , \nn && S = - \frac{1/2}{\beta} \sum_{i\omega } \frac{1}{L} \sum_{k} \sum_{p = 0}^{n} \frac{\frac{\partial \mu_{p}(\chi,\lambda;i\omega;H)}{\partial \lambda} + \Theta(p - n + 1) z \gamma_{k} \frac{\partial t_{p}(\chi,\lambda;i\omega;H)}{\partial \lambda}}{ \mu_{p}(\chi,\lambda;i\omega;H) + \Theta(p - n + 1) z t_{p}(\chi,\lambda;i\omega;H) \gamma_{k} } . \label{vevofoperator} \eqa
\end{widetext}
An essential point is the emergence of an additional summation $\sum_{p = 0}^{n} \cdot\cdot\cdot$, representing the step of the renormalization group analysis, where $n \rightarrow \infty$ corresponds to the IR limit. We claim that the additional index $p$ encodes the information of a bulk geometry with renormalization group equations (6).

In order to make the emergent geometry apparent, we perform the continuation of $\frac{\mu_{n}(\chi,\lambda;i\omega;H) - \mu_{n-1}(\chi,\lambda;i\omega;H)}{n - (n-1)} \longrightarrow \frac{d \mu(\chi,\lambda;i\omega,r;H)}{d (r/a)}$, where $a$ is introduced to keep a length scale. Then, Eq. (6) can be rewritten as follows \bqa && \frac{d \mu(\chi,\lambda;i\omega,r;H)}{d (r/a)} = - \frac{t^{2}(\chi,\lambda;i\omega,r;H)}{\mu(\chi,\lambda;i\omega,r;H)} , \nn && \frac{d t(\chi,\lambda;i\omega,r;H)}{d (r/a)} = - t(\chi,\lambda;i\omega,r;H) + \frac{t^{2}(\chi,\lambda;i\omega,r;H)}{\mu(\chi,\lambda;i\omega,r;H)} . \nn \label{beta functions} \eqa

It is straightforward to solve these coupled equations. Considering $y(\chi,\lambda;i\omega,r;H) \equiv \frac{t(\chi,\lambda;i\omega,r;H)}{\mu(\chi,\lambda;i\omega,r;H)}$, we obtain \bqa \frac{d \ln y(\chi,\lambda;i\omega,r;H)}{d (r/a)} &=& y^{2}(\chi,\lambda;i\omega,r;H) \nn &+& y(\chi,\lambda;i\omega,r;H) - 1 , \eqa where the solution is given by \bqa && \exp \Big( - \frac{r}{a} \Big) = \Big(\frac{y(\chi,\lambda;i\omega,r;H)}{y_{0}}\Big)^{y_{+} - y_{-}} \nn && \times \Big(\frac{y(\chi,\lambda;i\omega,r;H) - y_{+}}{y_{0} - y_{+}}\Big)^{y_{-}} \Big(\frac{y_{0} - y_{-}}{y(\chi,\lambda;i\omega,r;H) - y_{-}}\Big)^{y_{+}} \nn \eqa with $y_{0} \equiv y(\chi,\lambda;i\omega,r = 0;H) = \frac{(J/N_{s}) \chi}{i\omega - \lambda - g \sigma H}$ and $y_{\pm} = \frac{- 1 \pm \sqrt{5}}{2}$. As a result, we find \bqa && \mu(\chi,\lambda;i\omega,r;H) = - (i\omega - \lambda - g \sigma H) \nn && \times \exp\Big( - \frac{1}{a} \int_{0}^{r} d r' y^{2}(\chi,\lambda;i\omega,r';H) \Big) , \nn && t(\chi,\lambda;i\omega,r;H) = - (J/N_{s}) \chi \nn && \times \exp\Big( - \frac{r}{a} + \frac{1}{a} \int_{0}^{r} d r' y(\chi,\lambda;i\omega,r';H) \Big) . \eqa

The renormalization group equation (10) shows two unstable fixed points, given by $y(\chi,\lambda;i\omega,r \rightarrow \infty;H) = y_{\pm}$, where $\mu(\chi,\lambda;i\omega,r \rightarrow \infty;H) = 0$ and $t(\chi,\lambda;i\omega,r \rightarrow \infty;H) \rightarrow \infty$ for $y_{+}$ while $\mu(\chi,\lambda;i\omega,r \rightarrow \infty;H) = 0$ and $t(\chi,\lambda;i\omega,r \rightarrow \infty;H) = 0$ for $y_{-}$, respectively, and three stable fixed points, given by $y(\chi,\lambda;i\omega,r \rightarrow \infty;H) \rightarrow \pm \infty$ and $y(\chi,\lambda;i\omega,r \rightarrow \infty;H) = 0$, where $\mu(\chi,\lambda;i\omega,r \rightarrow \infty;H) = 0$ and $t(\chi,\lambda;i\omega,r \rightarrow \infty;H) \rightarrow \infty$ for $+ \infty$, and $\mu(\chi,\lambda;i\omega,r \rightarrow \infty;H) = 0$ and $t(\chi,\lambda;i\omega,r \rightarrow \infty;H) = 0$ for $- \infty$ while $\mu(\chi,\lambda;i\omega,r \rightarrow \infty;H) = \mu^{*} \equiv - (i\omega - \lambda - g \sigma H)$ and $t(\chi,\lambda;i\omega,r \rightarrow \infty;H) = 0$ for $0$, respectively \cite{Real_Space_RG_Comment}. In particular, the last stable fixed point coincides with the stable fixed point of the Ising model under an external magnetic field in the real-space renormalization group approach a la Kadanoff \cite{Stamech_Textbook}.

Inserting these solutions into the effective free energy in the continuum representation for the additional coordinate, we obtain \bqa && F(\chi,\lambda;\Lambda;H) = - \frac{1}{\beta} \ln \int D \Psi_{\sigma}(i \omega,x,r) \nn && \times \exp\Big[ - \sum_{i \omega} \int_{-\infty}^{\infty} d x \int_{0}^{\Lambda} \frac{d r}{a} e^{- \frac{1}{a} \int_{0}^{r} d r' y^{2}(\chi,\lambda;i\omega,r';H)} \nn && \times \Psi_{\sigma}^{\dagger}(i \omega,x,r) \Bigl\{ - (i \omega - \lambda - g \sigma H) - \Theta(r / a - \Lambda + \delta) \nn && \times z [(i \omega - \lambda - g \sigma H) y(\chi,\lambda;i\omega,r;H)] \cos (- i \partial_{x} ) \Bigr\} \nn && \times \Psi_{\sigma}(i \omega,x,r) \Big] + z \frac{J}{N_{s}} \chi^{2} - N_{s} S \lambda , \eqa where $\delta$ is an infinitesimal positive constant. It is clear that the information on the emergent curved space is encoded into $e^{- \frac{1}{a} \int_{0}^{r} d r' y^{2}(\chi,\lambda;i\omega,r';H)}$ in $\int_{0}^{\Lambda} \frac{d r}{a}$ and the $r-$dependent kinetic-energy term with $y(\chi,\lambda;i\omega,r;H)$. The limit of $\Lambda \rightarrow \infty$ completes the structure of the emergent geometry. We emphasize that this effective free energy is classical in spite of the presence of the $\Psi_{\sigma}(i \omega,x,r)$ integral since the resulting field theory is noninteracting essentially. Self-consistent equations for order parameter fields are reformulated with the emergent radial coordinate $r$ from this effective free energy, where the integral expression for the hopping order parameter $\chi \rightarrow \chi(r)$ of Eq. (8) can be translated into a differential equation with respect to $r$, not shown here. Thermodynamic properties are formulated as follows: The uniform spin susceptibility is given by $\chi_{s}(\Lambda;T) = - \frac{\partial^{2} F(\chi,\lambda;\Lambda;H)}{\partial H^{2}} \Big|_{H \rightarrow 0}$, for example. One can translate this integral expression into a differential equation for the radial coordinate, referred to as Callan-Symanzik equation \cite{Callan_Symanzik_Eq}.
%
%

When both hopping and pairing order parameters are taken into account, the hopping parameter should be generalized into the hopping matrix in the Nambu-spinor basis representation \cite{PH_SU2_Comment}. Then, not only renormalization group equations but also self-consistent equations for order parameter fields are expressed by differential equations of matrix fields, technically more involved and challenging.

Formally, it is straightforward to extract out the metric tensor from this effective free energy. Let us take into account the renormalization group flow equation
\bqa
\gamma^{\mu\nu} T_{\mu\nu} + \sum_{i=1}^{n} \beta_i \left< {\cal O}^i \right> = {\cal A} ,
\eqa
where ${\cal A}$ denotes an anomaly. The energy-momentum tensor is given by \bqa && T_{\mu\nu} = \Big\{ \frac{\delta \mathcal{L}[\Psi_{\sigma}(\tau,x,r),\partial_{\mu} \Psi_{\sigma}(\tau,x,r)]}{\delta [\partial_{\mu} \Psi_{\sigma}(\tau,x,r)]} [\partial_{\nu} \Psi_{\sigma}(\tau,x,r)] \nn && - \delta_{\mu\nu} \mathcal{L}[\Psi_{\sigma}(\tau,x,r),\partial_{\mu} \Psi_{\sigma}(\tau,x,r);\tau,x,r] \Big\}_{r/a=\Lambda} . \eqa The $\beta-$functions are \bqa && \beta_{t} = \left. \frac{d t[\chi(r),\lambda(r);i\omega,r;H]}{d (r/a)} \right|_{r/a=\Lambda} , \nn && \beta_{\mu} = \left. \frac{d \mu[\chi(r),\lambda(r);i\omega,r;H]}{d (r/a)} \right|_{r/a=\Lambda} , \eqa and vacuum expectation values are \bqa \langle \mathcal{O}_{t} \rangle &=& z \langle \Theta(r / a - \Lambda + \delta) \Psi_{\sigma}^{\dagger}(\tau,x,r) \cos (- i \partial_{x} ) \Psi_{\sigma}(\tau,x,r) \rangle \nn &=& z \chi(r) {\large |}_{r/a=\Lambda}, \nn \langle \mathcal{O}_{\mu} \rangle &=& \langle \Psi_{\sigma}^{\dagger}(\tau,x,r) \Psi_{\sigma}(\tau,x,r) \rangle =  N_{s} S {\large |}_{r/a=\Lambda} . \eqa
%
%
These information derived from the Wilsonian renormalization group approach at any energy scale $r$ determine the metric $\gamma_{\mu\nu}$ as a function of $r$. If the theory has no anomaly, this metric can be easily lifted into a three-dimensional metric in the normal coordinate \cite{deBoer:1999tgo}
\bqa
ds^2 = dr^2 + \gamma_{\mu\nu} (r) \ dx^\mu dx^\nu .
\eqa

A cautious person may ask the connection to the multiscale entanglement renormalization ansatz, in short, MERA \cite{MERA_Review}, regarded to be one promising direction how to represent a ground-state wave-function. An essential aspect in MERA is to construct the ground-state wave-function through the real-space renormalization group framework, realized by both disentangler and isometry, where the disentangler takes into account the unitary evolution of the system via an interaction Hamiltonian and the isometry rescales the system to turn it back to the original system. In order to clarify this connection, we recall how to express the single-particle wave-function in the path-integral representation \cite{QM_Path_Integral_Textbook} \bqa && \psi(x_{f},t_{f}) = \int D x_{i} G(x_{f}t_{f};x_{i}t_{i}) \psi(x_{i},t_{i}) \nn && = \int D x_{i} \int_{x_{i}}^{x_{f}} D x(t) e^{\frac{i}{\hbar} \int_{t_{i}}^{t_{f}} d t L[x(t),\dot{x}(t)]} \psi(x_{i},t_{i}) , \eqa where $\psi(x_{i},t_{i})$ evolves into $\psi(x_{f},t_{f})$ through $G(x_{f}t_{f};x_{i}t_{i})$, given by the path-integral representation. It is straightforward to generalize this concept into the case of the many-particle ground-state wave-function \bqa && \Psi[\Phi(\bm{x},t_{f})] = \int D \Phi(\bm{x},t_{i}) \int_{\Phi(\bm{x},t_{i})}^{\Phi(\bm{x},t_{f})} D \psi(\bm{x},t) \nn && \exp\Big( \frac{i}{\hbar} \int_{t_{i}}^{t_{f}} d t \int d^{d} \bm{x} \mathcal{L}[\psi(\bm{x},t)] \Big) \Psi[\Phi(\bm{x},t_{i})] , \eqa where the ground-state wave-function $\Psi[\Phi(\bm{x},t_{i})]$ evolves into $\Psi[\Phi(\bm{x},t_{f})]$ via the many-particle Green's function in the path-integral representation, given by \bqa && G[\Phi(\bm{x},t_{f});\Phi(\bm{x},t_{i})] \nn && = \int_{\Phi(\bm{x},t_{i})}^{\Phi(\bm{x},t_{f})} D \psi(\bm{x},t) \exp\Big( \frac{i}{\hbar} \int_{t_{i}}^{t_{f}} d t \int d^{d} \bm{x} \mathcal{L}[\psi(\bm{x},t)] \Big) . \nn \eqa The MERA represents this many-body Green's function, based on the real-space renormalization group framework. This is exactly what we have performed in the imaginary-time formulation with $t_{i} \rightarrow 0$ and $t_{f} \rightarrow \beta$, where the disentangler and the isometry correspond to the integration of $-$ sublattice fermion fields and the rescaling procedure, respectively, in the path-integral representation of the partition function.

As the geometrical construction of the entanglement entropy in the MERA representation has been argued to resemble the Ryu-Takayanagi formula \cite{Ryu_Takayanagi_Formula} given by the area of a minimal surface in the gravity description \cite{MERA_Entanglement_Entropy}, it is necessary to evaluate the entanglement entropy \cite{Entanglement_Entropy_Review} within our formulation in order to verify the structure of the emergent geometry. We would like to point out that the way of our construction for the emergent gravity is rather field theoretic: Given the step of the renormalization group procedure, identified with the radial coordinate in the continuation procedure, the partition function with renormalized coupling constants at such an energy scale appears to encode the geometrical information through repeating the renormalization group procedure. One can calculate the entanglement entropy with renormalized coupling constants at $r$. This should be compared with the Ryu-Takayanagi formula \cite{Ryu_Takayanagi_Formula} in the geometrical description. This is an urgent future task.

In conclusion, we derived a geometrical description for strongly coupled field theories in one dimension, combining the mean-field theory with the Wilson's real-space renormalization group analysis. Based on this demonstration, we claimed that the AdS$_{3}$/CFT$_{2}$ duality conjecture may be interpreted as Landau-Ginzburg-Wilson theory with renormalized interaction vertices, described by the additional length scale. Unfortunately, it is not straightforward to generalize the present theoretical framework into that in higher dimensions than one dimension. The reason is quite simple: It is not clear how to perform the real-space renormalization group analysis above one dimension since the integration procedure of ``sublattice" fermion fields does not preserve the lattice structure. In this respect the mathematical structure of AdS$_{d+2}$/CFT$_{d+1}$ duality conjecture still remains unclear within our approach even if the mean-field theory plus the real-space renormalization group analysis seems reasonable at least conceptually.

\section*{Acknowledgement}

This study was supported by the Ministry of Education, Science, and Technology (No. NRF-2015R1C1A1A01051629, No. 2011-0030046, and NRF-2013R1A1A2A10057490) of the National Research Foundation of Korea (NRF) and by TJ Park Science Fellowship of the POSCO TJ Park Foundation. This work was also supported by the POSTECH Basic Science Research Institute Grant (2015). C. Park was also supported by the Korea Ministry of Education, Science and Technology, Gyeongsangbuk-Do and Pohang City. We would like to appreciate fruitful discussions in the APCTP workshop on Delocalisation Transitions in Disordered Systems in 2015.


\begin{thebibliography}{9}
\bibitem{PH_SU2_Comment} The effective Hamiltonian is given by \bqa && H_{eff} = - \frac{J}{4} \sum_{\sigma, \sigma' = \pm} \sum_{i} \left(\begin{array}{cc} f_{i\sigma}^{\dagger} & \varepsilon_{\sigma\sigma'} f_{i\sigma'} \end{array} \right) \left(\begin{array}{cc} \chi_{ii+1}^{\dagger} & \Delta_{ii+1} \\ \Delta_{ii+1}^{\dagger} & - \chi_{ii+1} \end{array} \right) \nn && \times \left(\begin{array}{c} f_{i+1\sigma} \\ \varepsilon_{\sigma\sigma'} f_{i+1\sigma}^{\dagger} \end{array} \right) + \frac{J}{4} \sum_{i} (|\chi_{ii+1}|^{2} + |\Delta_{ii+1}|^{2}) , \nonumber \eqa invariant under the particle-hole SU(2) gauge transformation $\left(\begin{array}{c} f_{i\sigma}' \\ \varepsilon_{\sigma\sigma'} f_{i\sigma}^{' \dagger} \end{array} \right) = \bm{g}_{i} \left(\begin{array}{c} f_{i\sigma} \\ \varepsilon_{\sigma\sigma'} f_{i\sigma}^{\dagger} \end{array} \right)$ and $\left(\begin{array}{cc} \chi_{ii+1}^{'\dagger} & \Delta_{ii+1}' \\ \Delta_{ii+1}^{'\dagger} & - \chi_{ii+1}' \end{array} \right) = \bm{g}_{i} \left(\begin{array}{cc} \chi_{ii+1}^{\dagger} & \Delta_{ii+1} \\ \Delta_{ii+1}^{\dagger} & - \chi_{ii+1} \end{array} \right) \bm{g}_{i+1}^{\dagger}$, where $\bm{g}_{i}$ belongs to the SU(2) group.
\bibitem{Real_Space_RG_Comment} The emergence of an insulating stable fixed point seems to imply an interesting feature of the real-space renormalization group approach. The momentum-space renormalization group approach allows a metallic stable fixed point, where momentum is a good quantum number associated with emergent U$_{k}(1)$ symmetry at each momentum point $k$. On the other hand, the real-space renormalization group approach gives rise to the insulating stable fixed point, where position is a good quantum number related with emergent U$_{x}(1)$ symmetry at each position $x$. This feature looks general in our mean-field plus real-space renormalization group framework, which may be involved with the emergence of AdS$_{2}$ in the IR limit.
\end{thebibliography}
\end{document}